\documentclass[12pt]{article}

\usepackage{epsfig,amsmath,amssymb,graphics,graphicx} 

\newcommand{\be}{\begin{equation}}
\newcommand{\ee}{\end{equation}}
\newcommand{\bi}[1]{\vspace{-3mm} \bibitem{#1}}

\begin{document}
%%%%%%%%%%%%%%%%%%%%%%%%%%%%%%%%%%%%%%%%%%%%%%%%%%%%%%%%%%%%%%%%%%%%%%%%%%

\begin{center}
{\it Theoretical and Mathematical Physics 158 (2008) 179-195}
\vskip 5 mm

\vspace{7mm}

{\Large \bf Fractional Generalization of Quantum  
Markovian Master Equation} \\
\vspace{4mm}
{\large \bf Vasily E. Tarasov}\\
Skobeltsyn Institute of Nuclear Physics, \\
Moscow State University, 119991 Moscow, Russia \\
E-mail: tarasov@theory.sinp.msu.ru
\end{center}

\begin{abstract}
We prove a generalization of the quantum Markovian equation for observables.
In this generalized equation, we use superoperators that are fractional powers of 
completely dissipative superoperators.
We prove that the suggested superoperators are infinitesimal generators 
of completely positive semigroups and describe the properties 
of this semigroup.
We solve the proposed fractional quantum Markovian equation  
for the harmonic oscillator with linear friction.
A fractional power of the Markovian superoperator can be considered 
a parameter describing a measure of "screening" of the environment
of the quantum system:
the environmental influence on the system is absent for $\alpha=0$, 
the environment completely influences the system for $\alpha=1$, 
and we have a powerlike environmental influence for $0<\alpha<1$. 
\end{abstract}

%%%PACS 03.65.-w, 03.65.Ta, 03.67.Lx, 03.067-a

%\vskip 3 mm

% 03.65.-w quantum mechanics
% 03.65.Ta Foundations, theory of measurement, miscellaneous
% 03.67.-a quantum information
% 03.67.Lx quantum computation

%%%%%%%%%%%%%%%%%%%%%%%%%%%%%%%%%%%%%%%%%%%%%%%%%%%%%%%%%%%%%%%%%%%%%%%%%

\section{Introduction}

%%%%%%%%%%%%%%%%%%%%%%%%%%%%%%%%%%%%%%%%%%%%%%%%%%%%%%%%%%%%%%%%%%%%%%%%%

Fractional calculus appeared in 1695, when Leibniz described the derivative 
of order $\alpha=1/2$ \cite{OS,SKM,Ross}.
Derivatives and integrals of noninteger order were studied by 
Leibniz, Liouville, Grunwald, Letnikov and Riemann.
Many books have now been written about fractional calculus and 
fractional differential equations
\cite{OS,SKM,MR,Podlubny,KST,K,N}.
Derivatives and integrals of noninteger order and 
fractional integro-differential equations 
have found many applications in recent studies in physics
(see, e.g., \cite{Zaslavsky1,GM,WBG,H} and \cite{Zaslavsky2,MS,MK1,MK2}).

In quantum mechanics, 
observables are given by self-adjoint operators.
The dynamical description of a quantum system is given by superoperators.
A superoperator is a map that assigns one operator some other operator.

The motion of a system is naturally described in terms of the 
infinitesimal change of the system.
The equation for a quantum observable is called the Heisenberg equation. 
For Hamiltonian quantum systems, the infinitesimal superoperator 
is defined by some form of derivation.
A derivation is a linear map ${\cal L}$ 
that satisfies the Leibnitz rule
${\cal L}(AB)=({\cal L}A)B+ A({\cal L}B)$ for any operators $A$ and $B$.
A fractional derivative can be defined as the fractional power
of the derivative (see, e.g., \cite{IJM}).
It is known that the infinitesimal generator ${\cal L}=(1/i\hbar)[H, \ . \ ]$, 
which is used for Hamiltonian systems, is a derivation of quantum observables.
In \cite{Heis}, we regarded a fractional power ${\cal L}^{\alpha}$ 
of the derivative operator ${\cal L}=(1/i\hbar)[H, \ . \ ]$
as a fractional derivative on a set of observables. 
As a result, we obtain a fractional generalization of the Heisenberg equation,
which allows generalizing the notion of Hamiltonian quantum systems. 
We note that a fractional generalization of classical Hamiltonian systems 
was suggested in \cite{FracHam} (also see \cite{JPA2006}).
In the general case, quantum systems are non-Hamiltonian
and ${\cal L}$ is not a derivation operator.
For a wide class of quantum systems, the infinitesimal generator ${\cal L}$ is 
completely dissipative \cite{Kossakowski,Dav,IngKos,kn3}.
Therefore, it is interesting to consider a fractional generalization
of the equation of motion for non-Hamiltonian quantum systems using
a fractional power of a completely dissipative superoperator.

The most general change of state of a non-Hamiltonian quantum system
is a quantum operation \cite{Kr1,Kr2,Kr3,Kr4,Schu,JPA}.
A quantum operation for a quantum system can be described starting
from a unitary evolution of some closed Hamiltonian system if
the quantum system is a part of the closed system \cite{ALV,Weiss}.
But situations can arise where it is difficult or impossible 
to find a Hamiltonian system that includes the given quantum system.
As a result, the theory of non-Hamiltonian quantum systems 
can be considered a fundamental generalization 
of the quantum mechanics of Hamiltonian systems \cite{Kossakowski,Dav,IngKos,kn3}.
The quantum operations that describe the dynamics of non-Hamiltonian
systems can be regarded as real completely positive
trace-preserving superoperators on some operator space.
These superoperators form a completely positive semigroup.
The infinitesimal generator of this semigroup is completely dissipative. 
The problem of non-Hamiltonian dynamics 
is to obtain an explicit form for the infinitesimal generator, 
which is in turn connected with the problem of determining 
the most general explicit form of this superoperator.
This problem was investigated in \cite{K1,K2,Lind1}.  
Here, we consider superoperators that are fractional powers of 
completely dissipative superoperators.
We prove that the suggested superoperators are infinitesimal generators 
of completely positive semigroups.
The quantum Markovian equations with a completely dissipative superoperator
are the most general form of the Markovian master equation 
describing the nonunitary evolution of a density operator
that is trace preserving and completely positive. 
We consider a fractional generalization of the quantum Markovian equation,
which is solved for the harmonic oscillator with friction.
We can assume that other solutions and properties described in
\cite{Lind2,SS,ISSSS,ISS,N0,N1,N2,N3,TN2} 
can also be considered for fractional generalizations of the quantum Markovian equation
and the Gorini-Kossakowski-Sudarshan equation \cite{K1,K2}.

A fractional power of infinitesimal generator can be considered 
a parameter describing a measure of "screening" of the environment. 
Using the interaction representation of the quantum Markovian equation,
we consider a fractional power $\alpha$ 
of the non-Hamiltonian part of the infinitesimal generator.
We obtain the Heisenberg equation for Hamiltonian systems
in the limit as $\alpha \rightarrow 0$. 
In the case $\alpha=1$, we have the usual quantum Markovian equation.
As a result, we can distinguish the following cases: 
(1) absence of the environmental influence ($\alpha=0$), 
(2) complete environmental influence ($\alpha=1$), and
(3) powerlike screening of the environmental influence ($0<\alpha<1$). 
The physical interpretation of the fractional quantum Markovian equation 
can be connected with an existence of 
a powerlike "screening" of the environmental influence.
Quantum computations by quantum operations with mixed states 
(see, e.g., \cite{JPA}) can be controlled by this parameter.
We assume that there exist stationary states of open quantum systems 
\cite{Dav1,Lind2,Spohn,Spohn2,AH,ISS,TN2} 
that depend on the fractional parameter.
We note that it is possible to consider quantum dynamics with 
a low fractionality by a generalization of the
method proposed in \cite{TZ2} (also see \cite{TP,TG}).

In Section 2, we briefly review of superoperators
on an operator Hilbert space and quantum operations and introduce the notation.
In Section 3, we consider the fractional power of a superoperator.
In Section 4, we suggest a fractional generalization of the quantum Markovian equation.
In Section 5, we describe the properties of the fractional semigroup.
In Sections 6 and 7, we solve the fractional equations 
for the quantum harmonic oscillator with and without friction.

%%%%%%%%%%%%%%%%%%%%%%%%%%%%%%%%%%%%%%%%%%%%%%%%%%%%%%%%%%%%%%%%%%%%%%%%%%

\section{Superoperator and quantum operations}

%%%%%%%%%%%%%%%%%%%%%%%%%%%%%%%%%%%%%%%%%%%%%%%%%%%%%%%%%%%%%%%%%%%%%%%%%%

$ \quad \ $
Quantum theories essentially consist of two structures: 
a kinematic structure describing the initial  
states and observables of the system,
and a dynamical structure describing the change of 
these states and observables with time.
In quantum mechanics,  
the states and observables can be given by operators.
The dynamical description of the quantum system is given by 
a superoperator, which is
a map from a set of operators into itself.

Let ${\cal M}$ be an operator space. 
We let ${\cal M}^{*}$ denote the space dual to ${\cal M}$.
Hence, ${\cal M}^{*}$ is the set of all 
linear functionals on ${\cal M}$. 
The classic denotations for an element of ${\cal M}$ 
are $|B)$ and $B$.
The symbols $(A|$ and $\omega$ denote the elements of ${\cal M}^{*}$. 
By the Riesz-Frechet theorem, any linear continuous functional $\omega$
on an operator Hilbert space ${\cal M}$ has the form
$\omega(B)=(A|B)$ for all $B \in {\cal M}$,
where $|A)$ is an element in ${\cal M}$.
Therefore, the element $A$ can be considered not 
only an element $|A)$ of ${\cal M}$, but also 
an element $(A|$ of the dual space ${\cal M}^{*}$. 
The symbol $(A|B)$ for a value of 
the functional $(A|$ on the operator $|B)$ 
is the graphic combination of the symbols $(A|$ and $|B)$.  \\

\noindent {\large Definition 1.} 
{\it A linear superoperator is a map ${\cal L}$ 
from an operator space ${\cal M}$ into itself such that 
the relation
\[ {\cal L} (aA+bB) = a{\cal L} (A) + b{\cal L} (B) \]
is satisfied for all $A, B \in D ({\cal L}) \subset {\cal M} $, 
where $D({\cal L})$ is the domain of ${\cal L}$ 
and $a,b \in \mathbb{C}$. } \\

A superoperator ${\cal L}$ assigns each operator $A \in D ({\cal L})$ 
the operator ${\cal L}(A)$. \\

\noindent {\large Definition 2.} 
{\it  Let ${\cal L}$ be a superoperator on ${\cal M}$.
An adjoint superoperator of ${\cal L}$ 
is a superoperator $\Lambda=\bar {\cal L}$ on ${\cal M}^{*}$ 
such that 
\be \label{Lam}
(\Lambda (A) |B) = (A | {\cal L} (B)) \ee
for all $B \in D ({\cal L}) \subset {\cal M}$ 
and $A \in D(\Lambda) \subset {\cal M}^{*}$.} \\

Let ${\cal M}$ be an operator Hilbert space and
${\cal L}$ be a superoperator on ${\cal M}$. 
Then $(A|B)=Tr[A^{\dagger} B]$, and equation (\ref{Lam}) becomes
\[ Tr [(\Lambda (A))^{\dagger} B] =Tr [A^{\dagger} {\cal L} (B)] . \]

If ${\cal M}$ is an operator Hilbert space, 
then by the Riesz-Frechet theorem,  
${\cal M}$ and ${\cal M}^{*}$ are isomorphic, and
we can define the self-adjoint superoperators. \\

\noindent
{\large Definition 3.} 
{\it  A self-adjoint superoperator is a superoperator ${\cal L}$ 
on a Hilbert operator space ${\cal M}$ such that
$({\cal L} (A) |B) = (A | {\cal L} (B))$ for all 
$A, B \in D ({\cal L}) \subset {\cal M}$
and $D ({\cal L}) =D (\bar {\cal L})$.} \\

Let ${\cal M}$ be a normed operator space.
The superoperator ${\cal L}$ is said to be called bounded if
$\|{\cal L}(A)\|_{\cal M} \le c \|A\|_{\cal M}$ 
for some constant $c$ and all $A \in {\cal M}$.
The value 
\[ \|{\cal L}\|=\sup_{A \ne 0} 
\frac{\|{\cal L}(A)\|_{\cal M}}{\|A\|_{\cal M}}  \]
is called the norm of the superoperator ${\cal L}$.
If ${\cal M}$ is a normed space and 
${\cal L}$ is a bounded superoperator, then 
$\| \bar {\cal L} \| = \| {\cal L} \|$. 

In quantum theory, the class of 
real superoperators is the most important. \\

\noindent {\large Definition 4.} 
{\it  Let ${\cal M}$ be an operator space and 
$A^{\dagger}$ be an adjoint operator of $A\in {\cal M}$.
A  real superoperator is a superoperator 
${\cal L}$ on ${\cal M}$ such that
\[ [{\cal L} (A)]^{\dagger} = {\cal L} (A^{\dagger}) \] 
for all $A \in D ({\cal L}) \subset {\cal M}$ 
and $A^{\dagger}\in D ({\cal L})$.} \\

If ${\cal L}$ is a real superoperator, then $\Lambda=\bar {\cal L}$ is real.
If ${\cal L}$ is a real superoperator
and $A$ is a self-adjoint operator $A^{\dagger}=A \in D({\cal L})$, 
then the operator $B={\cal L}(A)$ is self-adjoint.
Then superoperators from a set of quantum observables ${\cal M}$ 
into itself should be real. 
All possible dynamics of quantum systems
must be described by a set of real superoperators.  \\

\noindent {\large Definition 5.} 
{\it A nonnegative superoperator is a map  
${\cal L}$ from ${\cal M}$ into ${\cal M}$, such that 
${\cal L} (A^2) \ge 0$
for all $A^2=A^{\dagger}A \in D ({\cal L}) \subset {\cal M}$.
A positive superoperator is a map 
${\cal L}$ from ${\cal M}$ into itself such that
${\cal L}$ is nonnegative and ${\cal L}(A)=0$ 
if and only if $A=0$.} \\

Let ${\cal M}$ denote an operator algebra.
A left superoperator corresponding to $A \in {\cal M}$ 
is a superoperator $L_A$ on ${\cal M}$ such that $L_A C=AC$ 
for all $C \in {\cal M}$. 
We can think of $L_A$ as meaning left multiplication by $A$. 
A right superoperator corresponding to $A \in {\cal M}$ 
is a superoperator $R_A$ on ${\cal M}$ such that 
$R_A C=CA$ for all $C \in {\cal M}$. 

The most general state change of a quantum system
is a called a { \it quantum operation} \cite{Kr1,Kr2,Kr3,Kr4,Schu,JPA}.
A quantum operation is described by a superoperator $\hat{\cal E}$
that is a map on a set of density operators.
If $\rho$ is a density operator, then $\hat{\cal E}(\rho)$
should also be a density operator.
Any density operator $\rho_t=\rho(t)$ is
a self-adjoint ($\rho^{\dagger}_{t}=\rho_{t}$),
positive ($\rho_{t}>0$) operator with unit trace ($ Tr\rho_{t}=1$).
Therefore, for a superoperator $\hat{\cal E}$
to be the quantum operation,  
the following conditions must be satisfied:
\begin{enumerate}
\item
The superoperator $\hat{\cal E}$ is a real superoperator, i.e.
$\Bigl(\hat{\cal E}(A)\Bigr)^{\dagger}=\hat{\cal E}(A^{\dagger})$
for all $A$.
The real superoperator $\hat{\cal E}$ maps the self-adjoint operator
$\rho$ to the self-adjoint operator $\hat{\cal E}(\rho)$:
\ $(\hat{\cal E}(\rho))^{\dagger}=\hat{\cal E}(\rho)$.
\item
The superoperator $\hat{\cal E}$ is a positive superoperator,
i.e. $\hat{\cal E}$ maps positive operators to positive operators:
\ $\hat{\cal E}(A^{2}) >0$ for all $A\not=0$ or
$\hat{\cal E}(\rho)\ge 0$.
\item
The superoperator $\hat{\cal E}$ is a trace-preserving map, i.e.
$(I|\hat{\cal E}|\rho)=(\hat{\cal E}^{\dagger}(I)|\rho)=1$
or $\hat{\cal E}^{\dagger}(I)=I$.
\end{enumerate}

Moreover, we assume that the superoperator $\hat{\cal E}$
is not only positive but also completely positive \cite{Arveson}.
The superoperator $\hat{\cal E}$ is a {\it  completely positive}
map from an operator space ${\cal M}$ into itself if 
\[ \sum^{n}_{k=1} \sum^{n}_{l=1} B^{\dagger}_{k}
\hat{\cal E}(A^{\dagger}_kA_l)B_l \ge 0 \]
for all operators $A_k$, $B_k \in {\cal M}$ and any integer $n$. \\

Let the superoperator $\hat{\cal E}$ be a convex linear map
on the set of density operators, i.e.
\[ \hat{\cal E}\Bigl(\sum_{s} \lambda_{s} \rho_{s}\Bigr)=
\sum_{s} \lambda_{s} \hat{\cal E}(\rho_{s}), \]
where $0<\lambda_{s}<1$ for all $s$, and $\sum_{s} \lambda_{s}=1$.
Any convex linear map of density operators
can be uniquely extended to a {\it  linear} map on self-adjoint operators.
We note that any linear completely positive superoperator can be represented by
\[ \hat{\cal E}=\sum^{m}_{k=1} \hat L_{A_{k}} \hat R_{A^{\dagger}_{k}}: \quad
\hat{\cal E}(\rho)=\sum^{m}_{k=1} A_k \rho A^{\dagger}_k. \]
If this superoperator is trace-preserving, then
\[ \sum^{m}_{k=1} A^{\dagger}_{k} A_{k}=I. \]

Because all processes occur in time, it is natural to consider 
quantum operations $\hat {\cal E}(t,t_{0})$
that depend on time. 
Let the linear superoperators $\hat {\cal E}(t,t_{0})$ form a completely
positive quantum semigroup \cite{AL} such that 
\be \label{qo1}
\frac{d}{dt}\hat{\cal E}(t,t_{0})= \hat\Lambda_t \hat{\cal E}(t,t_{0}), \ee
where $\hat \Lambda_t$ is an infinitesimal generator of 
the semigroup \cite{Lind1,AL,kn3}. 
The evolution of a density operator $\rho$ is described by
\[ \hat{\cal E}(t,t_{0}) \rho(t_0)= \rho(t) . \]
We consider quantum operations $\hat{\cal E}(t,t_0)$ 
with an infinitesimal generator $\hat \Lambda$ such that 
the adjoint superoperator $ {\cal L}$ is completely dissipative, i.e.
\[ {\cal L} (A_{k}A_{l})-
{\cal L}(A_{k}) A_{l}-A_{k} {\cal L}(A_{l}) \ge 0, \]
for all $A_1,...,A_n \in D({\cal L})$ such that $A_kA_l \in D({\cal L})$.
The superoperator ${\cal L}$ describes the dynamics of observables 
of a non-Hamiltonian quantum system.
The completely dissipative superoperators are infinitesimal generators 
of completely positive semigroups  
$\{\Phi_t | \ t>0\}$ that are adjoint to $\{ \hat{\cal E}_t | \ t>0\}$, 
where $\hat{\cal E}_t= \hat{\cal E}(t,0)$.

%%%%%%%%%%%%%%%%%%%%%%%%%%%%%%%%%%%%%%%%%%%%%%%%%%%%%%%%%%%%%%%%%%%%%%%%%

\section{Fractional power of a superoperator}

%%%%%%%%%%%%%%%%%%%%%%%%%%%%%%%%%%%%%%%%%%%%%%%%%%%%%%%%%%%%%%%%%%%%%%%%%%

$ \quad \ $ 
Let ${\cal L}$ be a closed linear superoperator
with an everywhere dense domain $D({\cal L})$
and a resolvent $R(z,{\cal L})$ on the negative semiaxis, and 
satisfy the condition
\be \label{Rcond}
\| R(-z,{\cal L}) \| \le M / z, \quad (z>0, M>0) . \ee
We note that
\[ R(-z,{\cal L})= (zL_I+{\cal L})^{-1} . \]
The superoperator
\be \label{LaA1}
{\cal L}^{\alpha}=\frac{\sin \pi \alpha }{\pi} 
\int^{\infty}_0 dz\, z^{\alpha-1} R(-z,{\cal L}) \, {\cal L}
\ee
is defined on $D({\cal L})$ for $0< \alpha <1$ and 
is called a fractional power of the superoperator ${\cal L}$ \cite{HP,Yosida}.
We note that the superoperator ${\cal L}^{\alpha}$ allows a closure.
If a closed superoperator ${\cal L}$ satisfies condition (\ref{Rcond}), 
then ${\cal L}^{\alpha} {\cal L}^{\beta}={\cal L}^{\alpha+\beta}$ 
for $\alpha, \beta>0$, and $\alpha+\beta<1$. 

Let ${\cal L}$ be a closed generating superoperator of the semigroup 
$\{\Phi_{t} | \ t \ge 0\}$.
Then the fractional power ${\cal L}^{\alpha}$ of ${\cal L}$ is given by
\[ {\cal L}^{\alpha}= 
\frac{1}{\Gamma(-\alpha)} \int^{\infty}_0 dz\, z^{-\alpha-1} (\Phi_z-L_I) , \]
which is called the {\it Balakrishnan formula}.

The resolvent for the superoperator ${\cal L}^{\alpha}$ 
can be found by the equation 
\[ R(-z,{\cal L}^{\alpha})=
(zL_I+ {\cal L}^{\alpha})^{-1}= \]
\[ =\frac{\sin \pi \alpha }{\pi} \int^{\infty}_0 dx\, 
\frac{x^{\alpha}}{z^2+2 zx^{\alpha} \cos \pi \alpha +x^{2 \alpha} } 
\, R(-x,{\cal L}) , \]
called {\it Kato's formula}. 
It follows from this formula that the inequality
\[ \| R(-z, {\cal L}^{\alpha}) \| \le M / z, \quad (z>0) ,  \]
is satisfied with the constant $M$ in inequality (\ref{Rcond})
for the superoperator ${\cal L}$. 
It follows from the inequality
\[ \|z R(-z,{\cal L}) \|=\|z(zL_I+{\cal L})^{-1}\| \le M \]
for all $z>0$ that the superoperator $z(zL_I+{\cal L})^{-1}$
is uniformly bounded in every sector of the complex plane 
given by the relation $|\arg z| \le \phi$ 
for $\phi$ not greater than some number $\pi - \psi$, ($0<\psi<\pi$).
Then the superoperator $zR(-z, {\cal L}^{\alpha})$
is uniformly bounded in every sector of the complex plane
such that $|\arg z| \le \phi$  for $\phi<\pi - \alpha \psi$.

Let ${\cal L}$ be a closed generating superoperator of 
the semigroup $\{\Phi_{t} | \ t \ge 0\}$.
Then the superoperators
\be \label{BPf}
\Phi^{(\alpha)}_t=\int^{\infty}_0 ds \, f_{\alpha}(t,s) \, \Phi_s , 
\quad (t>0) , \ee
form a semigroup such that
${\cal L}^{\alpha}$ is an infinitesimal generator of $\Phi^{(\alpha)}_t$. 
Equation (\ref{BPf}) is called the Bochner-Phillips formula. 

In equation (\ref{BPf}), we use the function
\be \label{fats}
f_{\alpha}(t,s)=\frac{1}{2\pi i} \int^{a+\i\infty}_{a-i\infty}
dz \, \exp (sz-tz^{\alpha}) ,
\ee
where $a,t>0$, $s \ge 0 $, and $0<\alpha <1$.
The branch of $z^{\alpha}$ is chosen such that $Re(z^{\alpha})>0$
for $Re(z)>0$.
This branch is a one-valued function in the $z$ plane 
cut along the negative real axis.
This integral obviously converges by virtue of the factor $\exp(-tz^{\alpha})$.
The function $f_{\alpha}(t,s)$ has the following properties: 

\begin{enumerate}
\item
For all $s>0$, the function $f_{\alpha}(t,s)$ is nonnegative:
$f_{\alpha}(t,s)\ge 0$. 
\item
We have the identity
\[ \int^{\infty}_0 ds \, f_{\alpha}(t,s) =1 . \]
\item
For $t>0$ and $x>0$, 
\[ \int^{\infty}_0 ds \, e^{-sx} \, f_{\alpha}(t,s) = e^{-tx^{\alpha}}. \]
\item
Passing from the integration contour in (\ref{fats})
to the contour crossing of the two rays $r\, \exp(-i \theta)$ and 
$r\, \exp(+i \theta)$, where $r \in (0,\infty)$, 
and $\pi/2 \le \theta \le \pi$, we obtain
\[
f_{\alpha}(t,s)=\frac{1}{\pi} \int^{\infty}_0  dr \,
\exp (sr \cos \theta - t r^{\alpha} \cos (\alpha \theta)) \cdot \]
\be \label{freal}
 \cdot \sin (sr \sin \theta - 
t r^{\alpha} \sin (\alpha \theta)+ \theta) . \ee
\item
If $\alpha=1/2$, then $\theta = \pi$, and
\[ f_{1/2}(t,s)=\frac{1}{\pi} \int^{\infty}_0 dr \,
e^{-sr} \sin ( t \sqrt{r} )=
\frac{t}{2 \sqrt{\pi} s^{3/2}} e^{-t^2/4s} . \]
which is a corollary of equation (\ref{freal}).
\end{enumerate}

%%%%%%%%%%%%%%%%%%%%%%%%%%%%%%%%%%%%%%%%%%%%%%%%%%%%%%%%%%%%%%%%%%%%%%%%%%

\section{Fractional quantum Markovian equation}

%%%%%%%%%%%%%%%%%%%%%%%%%%%%%%%%%%%%%%%%%%%%%%%%%%%%%%%%%%%%%%%%%%%%%%%%%%
%%%\subsection{quantum Markovian equation}
%%%%%%%%%%%%%%%%%%%%%%%%%%%%%%%%%%%%%%%%%%%%%%%%%%%%%%%%%%%%%%%%%%%%%%%%%%

$ \quad \ $
The motion of a systems is naturally described in terms of the
infinitesimal change.
This change can be described by an infinitesimal generator. 
One problem of the non-Hamiltonian dynamics 
is to obtain an explicit form 
of the infinitesimal generator. 
For this, it is necessary to find 
the most general explicit form of this superoperator.
The problem was investigated in \cite{K1,K2,Lind1}
for completely dissipative superoperators.  
Lindblad showed that there exists a one-to-one correspondence 
between the completely positive norm-continuous semigroups and 
completely dissipative generating superoperators \cite{Lind1}.
Lindblad's structural theorem gives 
the most general form of a completely dissipative superoperator. \\

\noindent {\large Theorem 1.} 
{\it  A generating superoperator ${\cal L}_V$ of a completely positive 
unity-preserving semigroup $\{\Phi_t=\exp(-t{\cal L}_V)| \ t \ge 0\}$
on an operator space ${\cal M}$ can be represented in the form 
\be \label{s153}
-{\cal L}_V (A) =- \frac{1}{i\hbar}[H, A] + 
\frac{1}{2 \hbar} \sum^{\infty}_{k=1} 
\Bigl(V^{\dagger}_{k}[A, V_{k}] + [V^{\dagger}_{k}, A] V_{k} \Bigr) ,
\ee
where $H$, $V_{k}$, $\sum_k V^{\dagger}_{k}, V^{\dagger}_{k} V_{k} \in {\cal M}$.} \\

We note that the form of ${\cal L}_V$ is not uniquely fixed by (\ref{s153}).  
Indeed, formula (\ref{s153}) preserves its form under the changes
\[ V_k \ \rightarrow \ V_k+ a_k I , \quad
H \ \rightarrow \ H+\frac{1}{2i\hbar} 
\sum^{\infty}_{k=1} (a^{*}_kV_k-a_kV^{\dagger}_k) , \]
where $a_k$ are arbitrary complex numbers. 

Using $A_t=\Phi_t(A)$, where $\Phi_t=\exp(-t{\cal L}_V)$, we obtain the equation 
\be \label{LindA1}
\frac{d}{dt} A_t=-\frac{1}{i\hbar}[H,A_t]+ 
\frac{1}{2 \hbar} \sum^{\infty}_{k=1} \Bigl(V^{\dagger}_k [A_t, V_k] +
[V^{\dagger}_k, A_t] V_k \Bigr) , \ee
where ${\cal L}_V$ is defined by (\ref{s153}). 
This is called the {\it quantum Markovian equation} for the observable $A$. 

The Lindblad theorem gives an explicit form of
the equations of motion if the following restrictions are satisfied 
(here $\Lambda_V$ is adjoint to ${\cal L}_V$): 

\begin{enumerate}
\item
${\cal L}_V$ and $\Lambda_V$ are bounded superoperators and 
\item 
${\cal L}_V$ and $\Lambda_V$ are completely dissipative superoperators. 
\end{enumerate}

Davies extended the Lindblad result 
to a class of quantum dynamical semigroups  
with unbounded generating superoperators \cite{Davies2}.

%%%%%%%%%%%%%%%%%%%%%%%%%%%%%%%%%%%%%%%%%%%%%%%%%%%%%%%%%%%%%%%%%%%%%%%%%%%

We consider quantum Markovian equation (\ref{LindA1})
for an observable $A_t$.
We rewrite this equation in the form
\be \label{Lind1a}
\frac{d}{dt} A_t= -{\cal L}_V A_t , \ee
where ${\cal L}_V$ denotes the Markovian superoperator
\be \label{Lind1b}
{\cal L}_V =L^{-}_{H}+ \frac{i}{2} \sum^{\infty}_{k=1} 
\Bigl(L_{V^{\dagger}_k} L^{-}_{V_k}-L^{-}_{V^{\dagger}_k} R_{V_k}  \Bigr) . \ee
Here, we use the superoperators of left multiplication $L_V$ and
right multiplication $R_V$ determined by the relations $L_VA=VA$ and $R_VA=AV$.
The superoperator $L^{-}_H$ is a left Lie multiplication by $A$ such that 
\be \label{Lminus}
L^{-}_HA=\frac{1}{i \hbar} [H,A] .  \ee
If all operators $V_k$ are equal to zero, then ${\cal L}_V=L^{-}_H$,
and equations (\ref{Lind1a}) and (\ref{Lind1b}) give 
the Heisenberg equations for a Hamiltonian system. 
In the general case, the quantum system is non-Hamiltonian \cite{kn3}.

We obtain a fractional generalization
of the quantum Markovian equation. 
For this, we define a fractional power for 
the Markovian superoperator ${\cal L}_V$ in the form
\be \label{LaA3}
-({\cal L}_V)^{\alpha}=\frac{\sin \pi \alpha }{\pi} 
\int^{\infty}_0 dz \, 
z^{\alpha-1} R(-z,{\cal L}_V) \, {\cal L}_V \quad (0< \alpha <1) .
\ee
The superoperator $({\cal L}_V)^{\alpha}$ is called
a {\it fractional power of the Markovian superoperator}.
We note that $({\cal L}_V)^{\alpha}({\cal L}_V)^{\beta}=
({\cal L}_V)^{\alpha+\beta}$ 
for $\alpha, \beta>0$, and $\alpha+\beta<1$. 
As a result, we obtain the equation
\be \label{Lind2} 
\frac{d}{dt} A_t=-({\cal L}_V)^{\alpha} A_t , \ee
where $t$, $H/\hbar$ and $V_k/ \sqrt{\hbar}$ 
are dimensionless variables.
We call this is the {\it fractional quantum Markovian equation}. 

If $V_k=0$, then equation (\ref{Lind2}) 
gives the fractional Heisenberg equation \cite{Heis} of the form
\be \label{Heis2b} 
\frac{d}{dt} A_t=-(L^{-}_H)^{\alpha} A_t . \ee
The superoperator $(L^{-}_H)^{\alpha}$ is 
a {\it fractional power of the left Lie superoperator} (\ref{Lminus}). 
We note that this equation cannot be represented in the form
\[ \frac{d}{dt} A_t=-L^{-}_{H_{new}} A_t=
\frac{i}{\hbar} [H_{new}, A_t] \]
with some operator $H_{new}$.
Therefore, quantum systems described by (\ref{Heis2b}) 
are not Hamiltonian systems.
These systems are called the {\it fractional Hamiltonian quantum systems} (FHQS). 
Usual Hamiltonian quantum systems can be considered a special case of FHQS. 
We note that a fractional generalization of classical Hamiltonian systems 
was suggested in \cite{FracHam,JPA2006}.

Using the operators
\[ A_U(t) = U(t) A_t U^{\dagger}(t) , \quad W_k(t) = U(t) V_k U^{\dagger}(t) , \]
where $U(t)= \exp \{ (1/i\hbar) H \}$, 
we can write the quantum Markovian equation in the form
\be \label{Lind-I}
\frac{d}{dt} A_U(t)= - \tilde{\cal L}_W A_U(t) . \ee
The superoperator
\be \label{Lind-W}
\tilde{\cal L}_W =\frac{i}{2} \sum^{\infty}_{k=1} 
\Bigl(L_{W^{\dagger}_k} L^{-}_{W_k}-L^{-}_{W^{\dagger}_k} R_{W_k}  \Bigr) \ee
describes the non-Hamiltonian part of the evolution.
Equation (\ref{Lind-I}) is the quantum Markovian equation 
in the interaction representation.
The fractional generalization of this equation is
\be \label{Lind-I2}
\frac{d}{dt} A_U(t)= - (\tilde{\cal L}_W)^{\alpha} A_U(t) . \ee
Equation (\ref{Lind-I2}) is the fractional quantum Markovian equation 
in the interaction representation.
The parameter $\alpha$ can be considered 
a measure of the influence of the environment.
For $\alpha=1$, we have quantum Markovian equation (\ref{Lind-I}).
In the limit as $\alpha \rightarrow 0$, we obtain the Heisenberg equation
for the  quantum observable $A_t$ of a Hamiltonian system.
As a result, we can consider the physical interpretation of equations 
with a fractional power of the Markovian superoperator 
an influence of the environment.
The following cases can be considered in quantum theory: 
(1) absence of the environmental influence ($\alpha=0$), 
(2) complete environmental influence ($\alpha=1$), and 
(3) powerlike screening of the environmental influence ($0<\alpha<1$). 
The physical interpretation of fractional equation (\ref{Lind-I2})
can be connected with an existence of a powerlike screening 
of the environmental influence on the system.

%%%%%%%%%%%%%%%%%%%%%%%%%%%%%%%%%%%%%%%%%%%%%%%%%%%%%%%%%%%%%%%%%%%%%%%%%%
\section{Fractional semigroup}

$ \quad \ $
If we consider the Cauchy problem for equation (\ref{Lind1a}) 
with the initial condition given at the time $t=0$ by $A_0$,
then its solution can be written in the form
$A_t=\Phi_t A_0$. 
The one-parameter superoperators $\Phi_t$, $t \ge 0$ 
have the properties
\[ \Phi_t \Phi_s=\Phi_{t+s}, \quad (t,s >0) , \quad \Phi_0=L_I . \] 
As a result, the superoperators $\Phi_t$ form a semigroup, 
and the superoperator ${\cal L}_V$ is a generating superoperator 
of the semigroup $\{\Phi_{t} | \ t\ge 0\}$.

We consider the Cauchy problem for 
fractional quantum Markovian equation (\ref{Lind2}) 
with the initial condition given by $A_0$.
Then its solution can be presented in the form
\[ A_t(\alpha)=\Phi^{(\alpha)}_t A_0, \]
where the superoperators $\Phi^{(\alpha)}_t$, $t>0$,  
form a semigroup, 
which we call the {\it fractional semigroup}.
The superoperator $-({\cal L}_V)^{\alpha}$ is a 
generating superoperator of 
the semigroup $\{\Phi^{(\alpha)}_t| \ t\ge 0\}$.
We consider some properties of the fractional semigroups
$\{\Phi^{(\alpha)}_t | \ t>0\}$. 

The superoperators $\Phi^{(\alpha)}_t$ can be constructed
in terms of $\Phi_t$ by Bochner-Phillips formula (\ref{BPf}), 
where $f_{\alpha}(t,s)$ is defined in (\ref{fats}). 
If $A_t$ is a solution of quantum Markovian equation (\ref{Lind1a}),
then formula (\ref{BPf}) gives the solution
\[ A_t(\alpha)=\int^{\infty}_0 ds \, f_{\alpha}(t,s) A_s , \quad (t>0) \]
of fractional quantum Markovian equation (\ref{Lind2}). 

A linear superoperator $\Phi^{(\alpha)}_t$ is completely positive if 
\[ \sum_{i,j} B_i \Phi^{(\alpha)}_t(A^{\dagger}_i A_j) B_j \ge 0  \]
for any $A_i, B_i \in {\cal M}$. 
The following theorem states that the fractional semigroup
$\{\Phi^{(\alpha)}_t| \ t>0\}$ is completely positive. \\

\noindent
{\large  Theorem 2.}
{\it  If $\{\Phi_{t} | \ t>0\}$ is a completely positive semigroup of 
superoperator $\Phi_t$ on ${\cal M}$, 
then the fractional superoperators $\Phi^{(\alpha)}_t$ 
form a completely positive semigroup $\{ \Phi^{(\alpha)}_t | \ t>0\}$ .} \\

\noindent {\it  Proof.}
Bochner-Phillips formula (\ref{BPf}) gives 
\[ \sum_{i,j} B_i \Phi^{(\alpha)}_t(A^{\dagger}_i A_j) B_j =
\int^{\infty}_0 ds \, f_{\alpha}(t,s) 
\sum_{i,j} B_i \Phi_s(A^{\dagger}_i A_j) B_j   \]
for $t>0$.
Using 
\[ \sum_{i,j} B_i \Phi_s(A^{\dagger}_i A_j) B_j \ge 0 , \quad
f_{\alpha}(t,s)\ge 0  \quad (s>0) , \]
we obtain
\[ \sum_{i,j} B_i \Phi^{(\alpha)}_t(A^{\dagger}_i A_j) B_j \ge 0 . \ \ \ \Box \]

\noindent
{\large  Corollary.}
{\it  If $\Phi_t$, $t>0$, is a nonnegative one-parameter superoperator, i.e.,
$\Phi_t (A) \ge 0$ for $A \ge 0$, then
the superoperator $\Phi^{(\alpha)}_t$ is nonnegative, i.e., 
$\Phi^{(\alpha)}_t (A) \ge 0 $ for $A\ge 0$. } \\

Using the Bochner-Phillips formula and the property
$f_{\alpha}(t,s) \ge 0$,  $s>0$, 
we can easily prove that the superoperator $\Phi^{(\alpha)}_t$
is nonnegative, 
if $\Phi_t$, $t>0$ is a nonnegative one-parameter superoperator. 
This corollary can also be proved by using 
$B_1=I$, $A_1=A$, and $A_i=B_i=0$ ($i=2,...$) 
in the proof of the theorem. \\

In quantum theory, the class of real superoperators is the most important. 
Let $A^{\dagger} \in {\cal M}^{*}$ be adjoint to $A\in {\cal M}$.
A {\it  real superoperator} is a superoperator 
$\Phi_t$ on ${\cal M}$, such that
$(\Phi_t A)^{\dagger} = \Phi_t (A^{\dagger})$
for all $A \in D (\Phi_t) \subset {\cal M}$. 
A quantum observable is a self-adjoint operator. 
If $\Phi_t$ is a real superoperator
and $A$ is a self-adjoint operator, $A^{\dagger}=A$, 
then the operator $A_t=\Phi_t A$ is self-adjoint, i.e., 
$(\Phi_t A)^{\dagger}=\Phi_t A$. 
Let ${\cal M}$ be a set of quantum observables. 
Then superoperators on ${\cal M}$ into ${\cal M}$ must be real because 
quantum dynamics, i.e., 
temporal evolutions of quantum observables,
must be described by real superoperators. \\

\noindent
{\large  Theorem 3.}
{\it  If $\Phi_t$ is a real superoperator, 
then the superoperator $\Phi^{(\alpha)}_t$ is also real. } \\

\noindent {\it Proof.}
The Bochner-Phillips formula gives
\[ (\Phi^{(\alpha)}_t A)^{\dagger} =
\int^{\infty}_0 ds f^{*}_{\alpha}(t,s) \, (\Phi_s A)^{\dagger} , 
\quad (t>0) . \]
Using (\ref{freal}), we can easily see 
that $f^{*}_{\alpha}(t,s)=f_{\alpha}(t,s)$ 
is a real-valued function.
Then $(\Phi_t A)^{\dagger}=\Phi_t A^{\dagger}$ leads to 
$(\Phi^{(\alpha)}_t A)^{\dagger} = \Phi^{(\alpha)}_t (A^{\dagger})$
for all $A \in D (\Phi^{(\alpha)}_t) \subset {\cal M}$. $\ \ \ \Box$ \\

%%%%%%
 
If $\Phi_t$ is a superoperator on a Hilbert operator space ${\cal M}$, 
then an {\it  adjoint superoperator} of $\Phi_t$ 
is a superoperator $\hat{\cal E}_t$ on ${\cal M}^{*}$ such that 
\be \label{TrPhi} (\hat{\cal E}_t (A) |B) = (A | \Phi_t (B)) \ee
for all $B \in D (\Phi_t) \subset {\cal M}$ 
and some $A \in {\cal M}^{*}$. 
Using the Bochner-Phillips formula, we obtain
the following theorem. \\

\noindent
{\large  Theorem 4.}
{\it  If $\hat{\cal E}_t$ is an adjoint superoperator of $\Phi_t$, 
then the superoperator
\[ \hat{\cal E}^{(\alpha)}_t =
\int^{\infty}_0 ds \, f_{\alpha}(t,s) \ \hat{\cal E}_s , 
\quad (t>0) , \]
is an adjoint superoperator of $\Phi^{(\alpha)}_t$.} \\

\noindent {\it Proof.}
Let $\hat{\cal E}_t$ be adjoint to $\Phi_t$, i.e.
equation (\ref{TrPhi}) is satisfied. 
Then
\[ ( \hat{\cal E}^{(\alpha)}_t A| B) =
\int^{\infty}_0 ds \, f_{\alpha}(t,s) (\hat{\cal E}_s A|B) = \]
\[ =\int^{\infty}_0 ds \, f_{\alpha}(t,s) 
(A| \Phi_s B) = (A| \Phi^{(\alpha)}_t B) . \ \ \ \Box \]

It is known that $\hat{\cal E}_t$ is a real superoperator if
$\Phi_t$ is real. 
Analogously, if $\Phi^{(\alpha)}_t$ is a real superoperator, then
$\hat{\cal E}^{(\alpha)}_t$ is real.

Let $\{ \hat{\cal E}_t | t>0\}$ be a completely positive semigroup  
such that the density operator $\rho_t=\hat{\cal E}_t \rho_0$
is described by
\be \label{Lindrho}
\frac{d}{dt} \rho_t=- \hat \Lambda_V \rho_t , \ee
where $\hat \Lambda_V$ is adjoint to 
the Markovian superoperator ${\cal L}_V$.
The superoperator $\hat \Lambda_V$ can be represented in the form
\[ \hat \Lambda_V \rho_t=-\frac{1}{i\hbar}[H,\rho_t]+
\frac{1}{\hbar} \sum^{\infty}_{k=1} \Bigl(V_k \rho_t V^{\dagger}_k - 
(\rho_t V^{\dagger}_k V_k+V^{\dagger}_k V_k \rho_t) \Bigr) . \]
We note that equation (\ref{Lindrho}) with $V_k=0$
gives the von Neumann equation
\[ \frac{d}{dt} \rho_t=\frac{1}{i\hbar}[H,\rho_t] .  \]
The semigroup $\{ \hat{\cal E}^{(\alpha)}_t | \ t>0\}$
describes the evolution of the density operator
$\rho_t(\alpha)=\hat{\cal E}^{(\alpha)}_t \rho_0$
by the fractional equation
\[ \frac{d}{dt} \rho_t(\alpha)=- (\hat \Lambda_V)^{\alpha} \rho_t(\alpha) . \]
This is the {\it  fractional quantum Markovian equation for the density operator}.
For $V_k=0$, this equation gives 
\[ \frac{d}{dt} \rho_t=-(-L^{-}_H)^{\alpha} \rho_t . \] 
which can be called the {\it  fractional von Neumann equation}.

%%%%%%%%%%%%%%%%%%%%%%%%%%%%%%%%%%%%%%%%%%%%%%%%%%%%%%%%%%%%%%%%%%%%%%%%%%
\section{Fractional equation for the harmonic oscillator}

$ \quad \ $ 
We consider a quantum harmonic oscillator such that
\be \label{oscHam}
H=\frac{1}{2m} P^2 +\frac{m\omega^2}{2} Q^2, \quad V_k=0 , \ee
where $t$ and $P$ are dimensionless variables.
Then equation (\ref{Lind2}) (also see (\ref{Heis2b})) describes a harmonic oscillator.
For $A=Q$ and $A=P$, equation (\ref{Lind2}) for $\alpha=1$ gives
\[ \frac{d}{dt} Q_t=\frac{1}{m} P_t, \quad 
\frac{d}{dt} P_t=-m \omega^2 Q_t . \]
The well-known solutions of these equations are
\[ Q_t=Q_0 \cos (\omega t) +\frac{1}{m \omega} P_0 \sin (\omega t) , \]
\be \label{osc1}
P_t=P_0 \cos (\omega t) - m \omega Q_0 \sin (\omega t) . \ee
Using these solutions and the Bochner-Phillips formula, 
we can obtain solutions of the fractional equations
\be \label{ex2}
\frac{d}{dt} Q_t=- (L^{-}_{H})^{\alpha} Q_t , \quad 
\frac{d}{dt} P_t=- (L^{-}_{H})^{\alpha} P_t , \ee
where $H$ is given by (\ref{oscHam}). 
The solutions of fractional equations (\ref{ex2}) have the forms
\[ Q_t(\alpha)=\Phi^{(\alpha)}_t Q_0=
\int^{\infty}_0 ds f_{\alpha}(t,s) Q_s , \]
\be \label{osc2}
P_t(\alpha)=\Phi^{(\alpha)}_t P_0=
\int^{\infty}_0 ds f_{\alpha}(t,s) P_s . \ee
Substituting (\ref{osc1}) in (\ref{osc2}) gives \cite{Heis} the equations
\be \label{Hsol2a}
Q_t=Q_0 C_{\alpha}(t) +\frac{1}{m \omega} P_0 S_{\alpha}(t) , \quad
P_t=P_0 C_{\alpha}(t) - m \omega Q_0 S_{\alpha}(t) , \ee
where
\[ C_{\alpha}(t)=\int^{\infty}_0 ds \, f_{\alpha}(t,s)\, \cos(\omega s) , 
\quad 
S_{\alpha}(t)=\int^{\infty}_0 ds \, f_{\alpha}(t,s)\, \sin(\omega s) . \]
Equations (\ref{Hsol2a}) describe solutions of 
fractional equations (\ref{ex2}) for the quantum harmonic oscillator. 
For $\alpha=1/2$, we have 
\[ C_{1/2}(t)=\frac{t}{2 \sqrt{\pi}} \int^{\infty}_0 ds \, 
\frac{\cos(\omega s)}{s^{3/2}} \, e^{-t^2/4s} , \]
\[ S_{1/2}(t)=\frac{t}{2 \sqrt{\pi}} \int^{\infty}_0 ds \, 
\frac{\sin(\omega s)}{s^{3/2}} \, e^{-t^2/4s} . \]
These functions can be represented in terms of the Macdonald function
(see Sec. 2.5.37.1 in \cite{Prudnikov}), 
which is also called the modified Bessel function of the third kind.

It is easy to obtain the expectations 
\[ <Q_t>=x_0 C_{\alpha}(t) +\frac{1}{m \omega} p_0 S_{\alpha}(t) , \]
\[ <P_t>=p_0 C_{\alpha}(t) - m \omega x_0 S_{\alpha}(t) , \]
and the dispersions 
\[ D_t(Q)=\frac{a^2}{2} C^2_{\alpha}(t) + 
\frac{\hbar^2}{2a^2 m^2 \omega^2} S^2_{\alpha}(t) , \]
\[ D_t(P)=\frac{\hbar^2}{2a^2}
C^2_{\alpha}(t) +\frac{a^2 m^2 \omega^2}{2} S^2_{\alpha}(t) . \]
Here, we use the coordinate representation and the pure state
\be \label{Psi0}
\Psi(x)=<x|\Psi>= \frac{1}{\sqrt{a\sqrt{\pi}}} \exp\Bigl\{-\frac{(x-x_0)^2}{2a} +
\frac{i}{\hbar} p_0x \Bigr\} . \ee
The expectation and dispersion are defined by usual.
%%%\[ <A_t>=Tr[|\Psi><\Psi|A_t] =<\Psi|A_t|\Psi> ,  \]
%%%\[ D_t(A)=<A^2_t>-<A_t>^2=<\Psi|A^2_t|\Psi>-<\Psi|A_t|\Psi>^2 . \]

%%%%%%%%%%%%%%%%%%%%%%%%%%%%%%%%%%%%%%%%%%%%%%%%%%%%%%%%%%%%%%%%%%%%%%%%%%
\section{Fractional quantum Markovian equation for the oscillator with friction}

$ \quad \ $ 
We consider the fractional quantum Markovian equation with $V_k \ne 0$.
The basic assumption is that the general form of a 
bounded completely dissipative superoperator 
given by the quantum Markovian equation
also holds for an unbounded completely 
dissipative superoperator ${\cal L}_V$.
Another condition imposed on the operators $H$ and $V_k$
is that they are functions of the operators $Q$ and $P$ 
such that the obtained model is exactly solvable \cite{Lind2,SS}
(also see \cite{ISSSS,ISS}).
We assume that $V_k=V_k(Q,P)$ are 
the first-degree polynomials in $Q$ and $P$,
and that $H=H(Q,P)$ is a second-degree polynomial in $Q$ and $P$.
These assumptions are analogous to those used in classical
dynamics when friction forces proportional to the velocity are considered.
Then $V_k$ and $H$ are given in the forms:
\be \label{v-h} 
H = \frac{1}{2m} P^2 + \frac{m \omega^2}{2} Q^2 + \frac{\mu}{2} (PQ+QP)  ,
\quad V_k=a_kP+b_kQ , 
\ee 
where $a_k$ and $b_k $, $k=1,2$, are complex numbers.
It is easy to obtain
\[ {\cal L}_V Q = \frac{1}{m} P + \mu Q - \lambda Q , \]
\[ {\cal L}_V P = -m \omega^2 Q - \mu P - \lambda P , \] 
where
\[ \lambda =Im\Bigl (\sum^{n=2}_{k=1} a_kb^*_k\Bigr) =
-Im \Bigl(\sum^{n=2}_{k=1} a^*_k b_k\Bigr) . \]
Using the matrices 
\[ A = 
\left (\begin{array}{c} Q \\ P \\
\end{array}
\right)  , \quad 
M = \left (\begin{array}{cc} 
\mu - \lambda & \frac{1}{m} \\
-m \omega^2 & - \mu - \lambda
\end{array} \right) , \]
we write the quantum Markovian equation for $A_t$ as
\be \label{Lineq} 
\frac{d}{dt} A_{t} =MA_{t} , \ee
where ${\cal L}_V A_t =MA_t$. 
The solution of (\ref{Lineq}) is 
\[ A_{t} = \Phi_t A_0 = 
\sum^{\infty}_{n=0} \frac{t^n}{n!}{\cal L}^n_V A_0 =
\sum^{\infty}_{n=0} \frac{t^n}{n!} M^n A_0 . \] 
The matrix $M$ can be represented in the form $M=N^{-1} FN$, 
where $F$ is a diagonal matrix. 
Let $\nu$ be a complex parameter such that
$\nu^2 = \mu^2 - \omega^2$. Then we have
\[ N =
\left (\begin{array}{cc} 
m \omega^2 & \mu + \nu \\ 
m \omega^2 & \mu - \nu
\end{array} \right), \quad
N^{-1} = \frac{1}{2m \omega^2 \nu}
\left (\begin{array}{cc} 
- (\mu - \nu) & \mu + \nu \\ 
m \omega^2 & -m \omega^2
\end{array} \right) , \]
\[ F = \left (
\begin{array}{cc}
- (\lambda + \nu) & 0 \\ 
0 & - (\lambda - \nu)
\end{array} \right) . \]
Taking
\[ \Phi_t = \sum^{\infty}_{n=0} \frac{t^n}{n!} M^n =
N^{-1} \left(\sum^{\infty}_{n=0} \frac{t^n}{n!} F^n\right) N , \]
into account, we obtain the superoperator $\Phi_t$ in the form
\[ \Phi_t=e^{tM} =N^{-1} e^{tF} N = \]
\[ =e^{-\lambda t}
\left (\begin{array}{cc} 
\cosh (\nu t) + (\mu/\nu) \sinh (\nu t) &
(1/m \nu) \sinh (\nu t) \\ 
- (m \omega^2/\nu) \sinh (\nu t) & 
\cosh(\nu t) - (\mu/\nu) \sinh (\nu t)
\end{array} \right) . \]
As a result, we obtain
\[ %%%\label{LindQ1}
Q_t=e^{-\lambda t}[\cosh (\nu t) + \frac{\mu}{\nu} \sinh (\nu t)] Q_0 +
\frac{1}{m \nu} e^{-\lambda t} \sinh (\nu t) P_0  , \]
\be \label{LindP1}
P_t = - \frac{m \omega^2}{\nu} e^{-\lambda t} \sinh (\nu t) Q_0 +
e^{-\lambda t}[\cosh (\nu t) - \frac{\mu}{\nu} \sinh (\nu t)] P_0 . \ee

The fractional quantum Markovian equations for $Q_t$ and $P_t$ are
\be \label{Lind3} 
\frac{d}{dt} Q_t=-({\cal L}_V)^{\alpha} Q_t , \quad
\frac{d}{dt} Q_t=-({\cal L}_V)^{\alpha} Q_t , \ee
where $t$ and $V_k/ \sqrt{\hbar}$ are dimensionless variables.
The solutions of these fractional equations
are given by the Bochner-Phillips formula,  
\[ %%%\label{LindQ2}
Q_t(\alpha)=\Phi^{(\alpha)}_t Q_0=
\int^{\infty}_0 ds f_{\alpha}(t,s) Q_s , 
\quad (t>0) , \]
\be \label{LindP2}
P_t(\alpha)=\Phi^{(\alpha)}_t P_0=
\int^{\infty}_0 ds f_{\alpha}(t,s) P_s , 
\quad (t>0) , \ee
where $Q_s$ and $P_s$ are given by (\ref{LindP1}) and 
the function $f_{\alpha}(t,s)$ is defined in (\ref{fats}). 
Substituting (\ref{LindP1}) in (\ref{LindP2}) gives
\[ %%%\label{LindQ3}
Q_t(\alpha)=[Ch_{\alpha}(t) + \frac{\mu}{\nu} Sh_{\alpha}(t)] Q_0 +
\frac{1}{m \nu} Ch_{\alpha}(t) P_0  , \]
\be \label{LindP3}
P_t(\alpha) = - \frac{m \omega^2}{\nu} Sh_{\alpha}(t) Q_0 +
[Ch_{\alpha}(t) - \frac{\mu}{\nu} Sh_{\alpha}(t)] P_0 , \ee
where
\[ Ch_{\alpha}(t)=\int^{\infty}_0 ds \, f_{\alpha}(t,s)\, 
e^{-\lambda s} \cosh (\nu s) , \]
\[ Sh_{\alpha}(t)=\int^{\infty}_0 ds \, f_{\alpha}(t,s)\, 
e^{-\lambda s} \sinh (\nu s) . \]
For $\alpha=1/2$, we have
\[ Ch_{1/2}(t)=\frac{t}{2 \sqrt{\pi}} \int^{\infty}_0 ds \, 
\frac{\cosh(\nu s)}{s^{3/2}} \, e^{-t^2/4s - \lambda s} , \]
\[ Sh_{1/2}(t)=\frac{t}{2 \sqrt{\pi}} \int^{\infty}_0 ds \, 
\frac{\sinh (\nu s)}{s^{3/2}} \, e^{-t^2/4s-\lambda s} . \]
These functions can be represented in terms of the Macdonald function
(see Sec. 2.4.17.2 in \cite{Prudnikov}) such that
\[ Ch_{1/2}(t)=\frac{t}{2 \sqrt{\pi}} \Bigl[
V(t,\lambda,-\nu)+V(t,\lambda,\nu) \Bigr] , \]
\[ Sh_{1/2}(t)=\frac{t}{2 \sqrt{\pi}} \Bigl[
V(t,\lambda,-\nu)-V(t,\lambda,\nu) \Bigr] , \]
where we use the notation
\[ V(t,\lambda,\nu)=\Bigl(\frac{t^2+4\nu}{4 \lambda}\Bigr)^{1/4} 
K_{-1/2} \Bigl( 2 \sqrt{\frac{\lambda(t^2+4\nu)}{4}} \Bigr) , \]
where $Re(t^2)> Re(\nu)$, $Re(\lambda)>0$,
and $K_{\alpha}(z)$ is the Macdonald function \cite{OS,SKM}. 
%%%which is also called the modified Bessel function of the third kind.

As a result, equations (\ref{LindP3}) define a solution
of the fractional quantum Markovian equation for 
the harmonic oscillator with friction.

\section{Conclusion}

Quantum dynamics can be described by superoperators.
A map assigning each operator exactly one operator is called a superoperator.
It is natural to describe motion in terms of the 
infinitesimal change of a system.
The equation of motion for a quantum observable
is called the Heisenberg equation. 
For Hamiltonian quantum systems, the infinitesimal superoperator 
is some form of derivation. 
A linear map ${\cal L}$ satisfying the Leibnitz rule
${\cal L}(AB)=({\cal L}A)B+ A({\cal L}B)$ for all operators $A$ and $B$
is called derivation.
It is known that the infinitesimal generator ${\cal L}=(1/i\hbar)[H, \ . \ ]$, 
which is used for Hamiltonian systems, is a derivation of quantum observables.
We can regard a fractional power ${\cal L}^{\alpha}$ of the derivative 
${\cal L}=(1/i\hbar)[H, \ . \ ]$ as  a fractional derivative on
a set of quantum observables \cite{Heis}. 
As a result, we obtain a fractional generalization of 
the Heisenberg equation \cite{Heis}, which allows generalizing 
the notion of Hamiltonian quantum systems. 
In the general case, quantum systems are non-Hamiltonian
and ${\cal L}$ is not a derivation.
For a wide class of quantum systems, the infinitesimal generator ${\cal L}$ is 
completely dissipative \cite{Kossakowski,Dav,IngKos,kn3}.

Here, we consider a fractional generalization of 
the equation of motion for non-Hamiltonian quantum systems using
a fractional power of a completely dissipative superoperator.
We suggested a generalization of the quantum Markovian equation 
for quantum observables.
In this equation, we used a superoperator that is a fractional power of 
a completely dissipative superoperator.
We proved that the suggested superoperator 
is an infinitesimal generator of a completely positive semigroup and 
described properties of this semigroup.
We solved the proposed fractional quantum Markovian equation exactly  
for the harmonic oscillator with linear friction.
A fractional power $\alpha$ of the quantum Markovian superoperator 
can be considered a parameter 
describing a measure of "screening" of the environment.
We can separate the cases where $\alpha=0$,  
absence of the environmental influence;
where $\alpha=1$, complete environmental influence; 
and where $0<\alpha<1$, a powerlike environmental influence. 
A one-parameter description of a screening of the coupling between
the quantum system and the environment is thus 
a physical interpretation of a fractional power of 
the quantum Markovian superoperator.

We note that the quantum Markovian equation describes
a coupling between a quantum system and an environment (see \cite{ALV}).
Another physical interpretation of a fractional power of 
the infinitesimal generator is connected with 
Bochner-Phillips formula (\ref{BPf}) as follows.
Using the properties
\[ \int^{\infty}_0 f_{\alpha}(t,s)=1 , \quad \quad f_{\alpha}(t,s) \ge 0 \quad 
(for \ all \ \ s>0) , \]
we can assume that $f_{\alpha}(t,s)$ is the density of a probability distribution.
Then Bochner-Phillips formula (\ref{BPf}) can be considered 
a smoothing of the evolution $\Phi_t$ with respect to the time $s>0$.
This smoothing can be considered a screening of 
the environment of the quantum system. 

The function $f_{\alpha}(t,s)$
can be represented as the Levy distribution using a reparametrization. 
We note that Levy distributions are solutions of fractional 
equations (see, e.g., \cite{Zaslavsky2,SZ,CNSNS2008-1,Y})
that describe anomalous diffusion.
It is known that quantum Markovian equations are used to describe
the Brownian motion of quantum systems \cite{Lind2}.
Perhaps, the fractional generalization of 
quantum Markovian equations can be used to describe 
anomalous processes and random walks 
\cite{Zaslavsky2,MS,MK1,MK2} in quantum systems.

%%%%%%%%%%%%%%%%%%%%%%%%%%%%%%%%%%%%%%%%%%%%%%%%%%%%%%%%%%%%%%%%%%%%%%%%

%%%%%%%%%%%%%%%%%%%%%%%%%%%%%%%%%%%%%%%%%%%%%%%%%%%%%%%%%%%%%%%%%%%%%%%%%%
\end{document}